\documentclass[aps,prl,reprint,superscriptaddress]{revtex4-1}

\usepackage[dvips]{epsfig}

\usepackage[sort&compress]{natbib}

\begin{document}

\title{Background-free quantum frequency conversion of single photons from a semiconductor quantum dot}

\author{Serkan Ates} \email{serkan.ates@nist.gov}
\affiliation{Center for Nanoscale Science and Technology,
National Institute of Standards and Technology, Gaithersburg, MD
20899, USA}
\affiliation{Maryland NanoCenter, University of Maryland, College
Park, MD}
\author{Imad Agha}
\affiliation{Center for Nanoscale Science and Technology,
National Institute of Standards and Technology, Gaithersburg, MD
20899, USA}
\affiliation{Maryland NanoCenter, University of Maryland, College
Park, MD}
\author{Antonio Badolato}
\affiliation{Department of Physics and Astronomy, University of
Rochester, Rochester, New York 14627, USA}
\author{Kartik Srinivasan}\email{kartik.srinivasan@nist.gov}
\affiliation{Center for Nanoscale Science and Technology, National
Institute of Standards and Technology, Gaithersburg, MD 20899, USA}
\date{\today}

\date{\today}

\begin{abstract}
We demonstrate background-free quantum frequency conversion of single photons from an epitaxially-grown InAs quantum dot. Single photons at $\approx$\,980\,nm are combined with a pump laser near 1550\,nm inside a periodically-poled lithium niobate (PPLN) waveguide, generating single photons at $\approx$\,600\,nm. The large red-detuning between the pump and signal wavelengths ensures nearly background-free conversion, avoiding processes such as upconversion of anti-Stokes Raman scattered pump photons in the PPLN crystal. Second-order correlation measurements on the single photon stream are performed both before and after conversion, confirming the preservation of photon statistics during the frequency conversion process.
\end{abstract}



\maketitle

Quantum frequency conversion~\cite{ref:Kumar} is a useful resource
in interfacing quantum systems that can be connected by photons but
which operate at disparate frequencies.  It has been enabled by the
development of high-efficiency frequency conversion techniques in
$\chi^{(2)}$~\cite{ref:Fejer_IEEE} and $\chi^{(3)}$
materials~\cite{ref:Uesaka_Kazovksy,ref:Gnauck}, and been
demonstrated in experiments in which the quantum state of a light
field was shown to be preserved during the
process~\cite{ref:Huang_Kumar_PRL,ref:Kim_PRL,ref:Tanzili_Zbinden,ref:Rakher_NPhot_2010,ref:McGuinnes_PRL10,ref:Ikuta_Imoto,ref:Zaske}.
Other experiments have focused on the benefits associated
with shifting the frequency of light to a wavelength band in which
high-efficiency detectors
exist~\cite{ref:Vandevender_Kwiat_JMO,ref:Langrock_Fejer,ref:Albota_Wong_upconversion,ref:Thew_upconversion,ref:Ma_upconversion}.
Ideally, quantum frequency conversion should be background-free and
avoid the generation of noise photons that are spectrally
unresolvable from the frequency-converted quantum state. Sum- and
difference-frequency generation in $\chi^{(2)}$
materials~\cite{ref:Kumar} and four-wave-mixing Bragg scattering in
$\chi^{(3)}$ materials~\cite{ref:McKinstrie} can be, in principle,
background-free, meaning that signal photons are directly converted
to idler photons without amplification of vacuum fluctuations, which
can occur in processes such as degenerate
four-wave-mixing~\cite{ref:McKinstrie}. However, other processes,
such as broadband Raman scattering of pump photons, may still be a
source of noise. Indeed, frequency conversion of Raman noise photons
is a major noise source in $\chi^{(2)}$ systems such
as quasi-phase-matched periodically-poled lithium niobate (PPLN)
waveguides~\cite{ref:Langrock_Fejer}, and was the dominant noise
source in a recent demonstration of frequency conversion of
single photon Fock states~\cite{ref:Rakher_NPhot_2010},
limiting the purity of the frequency-converted single photon source.

Here, we demonstrate nearly background-free quantum frequency
conversion of single photons from a semiconductor quantum dot. We
work with quantum dots  emitting in the well-studied 900\,nm to
1000\,nm wavelength
range~\cite{ref:Michler,ref:Santori2,ref:Moreau,ref:Shields_NPhot},
and convert their single photon emission to 600 nm. By using a much
wider wavelength separation between the signal and pump photons, we
show that the signal-to-background level improves by about two
orders of magnitude with respect to Ref.
~\onlinecite{ref:Rakher_NPhot_2010}. Measurements of the photon
statistics before and after conversion indicate essentially no
degradation in purity of the single photon stream due to the
frequency conversion process.

\begin{figure}
\centerline{\includegraphics[width=8.5 cm]{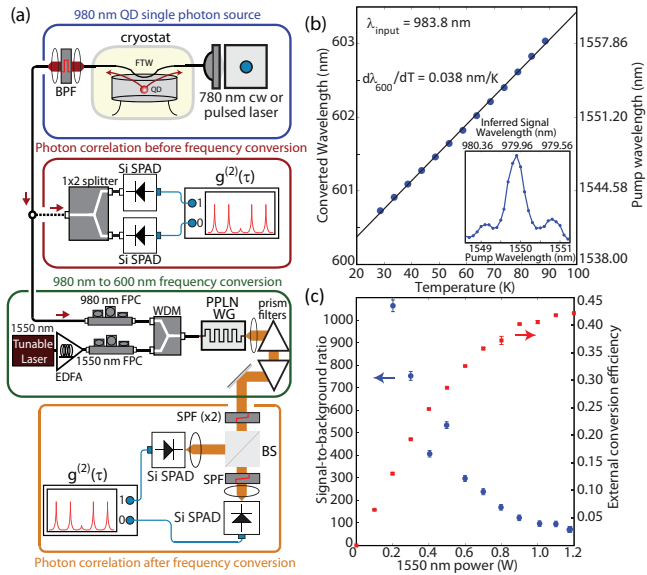}}
\caption{(a) Schematic of the experimental setups used within this
work. (b) Plot of the 600\,nm band wavelength as a function of PPLN
waveguide temperature, for a fixed 983.8\,nm signal and 1550\,nm
pump wavelength adjusted to maximize the converted signal. The inset
shows the quasi-phase-matching response of the PPLN waveguide, with
the signal fixed at 980\,nm and 1550\,nm pump wavelength varied. (c)
Signal-to-background ratio (left y-axis, blue points) and external
conversion efficiency (right y-axis, red points) as a function of
1550\,nm pump power coupled into the 980\,nm\,/1550\,nm WDM. The
external conversion efficiency includes all losses in the system.
The overall detection efficiency of the frequency conversion system
is given as the product of the external conversion efficiency and
the detector quantum efficiency
($\approx$\,67~$\%$)~\cite{ref:Microdisk_N1}.} \label{fig:Fig1}
\end{figure}

The experimental system we use for quantum frequency conversion
experiments is depicted in Fig.~\ref{fig:Fig1}(a). Our source of
single photons is a single InAs quantum dot (QD) in a fiber-coupled,
GaAs microdisk optical cavity~\cite{ref:Srinivasan16}, which is
excited above the GaAs band-edge by a continuous wave (cw) or pulsed
(50\,MHz repetition rate, 50\,ps pulse width) 780\,nm fiber-coupled
laser diode. While our previous work~\cite{ref:Rakher_NPhot_2010}
was focused on telecommunications-band (1300\,nm) to visible band
(710\,nm) wavelength conversion, here our QD emits photons at
$\approx$\,980\,nm. Emission from the QD is out-coupled by a fiber
taper waveguide (FTW) and spectrally isolated using a
$\approx$\,0.2~nm bandwidth volume reflective Bragg grating whose
input and output is coupled to single mode optical fiber. At this
point, we can either perform photon correlation measurements on the
980\,nm signal using a standard Hanbury-Brown and Twiss (HBT)
setup~\cite{ref:Michler}, or send it to the frequency conversion
setup. In the frequency conversion setup, we use a wavelength
division multiplexer (WDM) to combine the spectrally-filtered QD
emission with a strong (few hundred mW) 1550\,nm pump signal
that is generated by an external cavity tunable diode laser and
erbium-doped fiber amplifier (EDFA). Fiber polarization controllers
(FPCs) are used to adjust the polarization state of both the 980\,nm
and 1550\,nm beams. The combined signal and pump are coupled into a
2\,cm long, 5~$\%$ MgO-doped PPLN waveguide whose temperature can be
adjusted between 25~$^{\circ}$C and 90~$^{\circ}$C with
0.1~$^{\circ}$C resolution. Light is coupled into the waveguide
through a cleaved single mode optical fiber that is controlled by a
3-axis open-loop piezo stage and has a mode-field diameter of
5.8\,$\mu$m at 980\,nm. The coupling is optimized for the 980\,nm
band signal, at the expense of the 1550\,nm pump (additional pump
power compensates for the 1550\,nm coupling inefficiency). Light
exiting the PPLN waveguide is collimated and sent through two
dispersive prisms and two 750\,nm short pass edge filters (SPFs) to
eliminate residual 1550\,nm pump photons and frequency doubled
775\,nm pump photons from the signal. A second HBT setup is used to
measure photon statistics of the frequency converted signal.

We assess the properties of the frequency conversion setup by using
a narrow ($<$5\,MHz) linewidth 980\,nm band laser attenuated to a
power level of $\approx$\,30\,fW, similar to the expected average
power levels of our QD single photon sources. First, we measure the
quasi-phase-matching bandwidth of the PPLN waveguide.  The
temperature of the PPLN waveguide is set to 58.8~$^{\circ}$C, and
the 1550\,nm pump power is set to $\approx$\,800\,mW, close to the
value at which we achieve optimal conversion efficiency (described
further below). We then scan the 1550\,nm pump wavelength while
monitoring the frequency converted 600\,nm band signal on a silicon
single-photon avalanche diode (SPAD), as shown in the inset to Fig.~\ref{fig:Fig1}(b). The curve
approximately follows the theoretically expected sinc$^2$
response~\cite{ref:Fejer_IEEE}, and the inferred bandwidth in the
980\,nm band (determined by energy conservation) is
$\approx$\,0.20\,nm. We next repeat this measurement while keeping
the signal wavelength fixed at 983.8\,nm and varying the PPLN
waveguide temperature. The resulting plot of frequency converted
wavelength in the 600\,nm band against PPLN waveguide temperature is
shown in Fig.~\ref{fig:Fig1}(b), and indicates that the output
emission can be tuned by $\approx$\,2\,nm. This tuning would be
needed for precise spectral matching of the frequency-converted
single photons with a resonant quantum system operating near
600\,nm. We also perform experiments in which the 980\,nm signal
wavelength is varied, and the PPLN waveguide temperature and
1550\,nm band pump wavelength are adjusted to achieve
phase-matching. We observe that we can efficiently
($>35~\%$ external conversion efficiency) convert signals
in the wavelength region between 970\,nm and $>$\,995\,nm (the upper
wavelength limit of our laser), covering nearly the entire s-shell
emission range of the QD ensemble. This means that QDs emitting at
different wavelengths (unavoidable due to size dispersion during
growth) can be frequency converted to the same wavelength.

\begin{figure}[t]
\centerline{\includegraphics[width=8.5 cm]{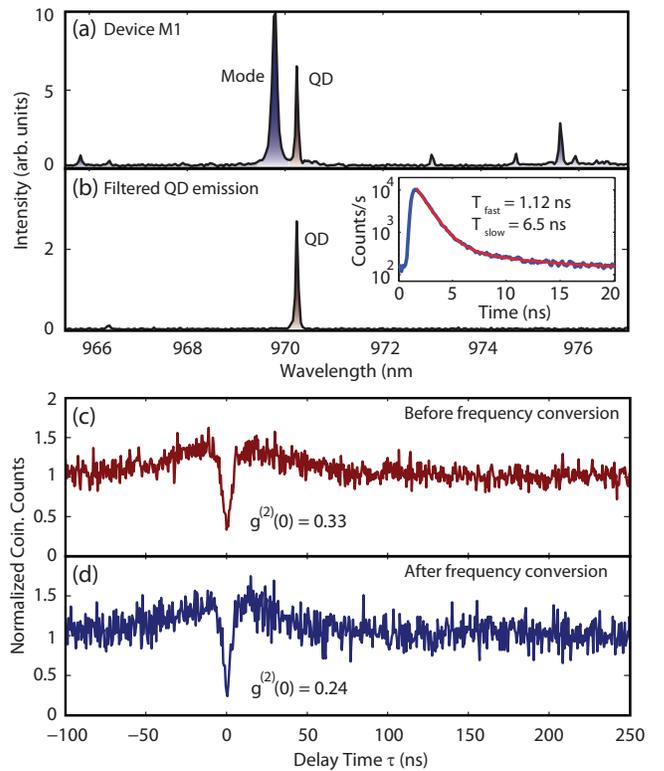}}
\caption{(a) Low-temperature $\mu$-PL spectrum of device M1 under
above-band cw excitation ($\lambda_{Exc.}\,=\,780$\,nm), showing a
single QD emission and a nearby cavity mode (Q\,=\,12500) around
970\,nm. (b) Spectrum of QD emission filtered by a volume Bragg
grating. (c)-(d) Second-order autocorrelation function measurements
performed on the QD emission line before and after the frequency
conversion resulted in $g^{(2)}(0)\,=\,0.33\,\pm\,0.03$ and
$g^{(2)}(0)\,=\,0.24\,\pm\,0.04$, respectively. inset: Time-resolved
PL of the QD emission revealed
$T_{fast}\,=\,1.12\,$ns$\,\pm\,0.05$\,ns and
$T_{slow}\,=\,6.5\,$ns$\,\pm\,0.1$\,ns.} \label{fig:Fig2}
\end{figure}

In our previous work~\cite{ref:Rakher_NPhot_2010}, in which the
signal was at 1300\,nm and pump at 1550\,nm, frequency conversion of
anti-Stokes photons generated by Raman scattering of the 1550\,nm
pump was thought to be the dominant source of noise, limiting the
achievable signal-to-background levels to about 7:1 (and as low as
2:1 for the highest conversion efficiencies). Use of a pulsed pump at 1550~nm~\cite{ref:Rakher_PRL_2011}
removes background emission that is temporally distinguishable from the single photon emission,
but does not improve the signal-to-background level. While better spectral
filtering does improve this ($>$~10:1 signal-to-background levels
were reported in a recent single photon downconverison
experiment~\cite{ref:Zaske}), it is perhaps more desirable to
suppress the source of the noise, for example, by increasing the
separation between the signal and red-detuned
pump~\cite{ref:Langrock_Fejer,ref:Pelc,ref:Dong_upconversion}. Here,
our pump-signal separation is nearly 600\,nm, suggesting that the
signal-to-background levels might be significantly improved. To test
this, we measure (Fig.~\ref{fig:Fig1}(c)) the signal-to-background
level as a function of 1550\,nm band pump power, determined by
comparing the ratio of the detected counts on the SPAD with and
without the presence of the 980\,nm band signal (and after
subtraction of the SPAD dark count rate of $\approx$~50~s$^{-1}$).
On the same graph, we plot the external conversion efficiency of the
system, which includes all PPLN waveguide input/output coupling,
free-space transmission, and spectral filtering losses (detector
quantum efficiency is not included). At the lowest 1550\,nm pump
powers (measured at the input of the 980\,nm/1550\,nm WDM), for
which the conversion efficiency is just a few percent, the
signal-to-background level exceeds 1000. As the pump power
increases, the signal-to-background level decreases, but still
remains above 100 for all but the highest 1550\,nm pump powers,
where the conversion efficiency has begun to roll off. For the
experiments that follow, we operate with a 35~$\%$ to 40~$\%$
external conversion efficiency, slightly higher than that used in
ref.~\onlinecite{ref:Rakher_NPhot_2010}, and a signal-to-background
level $>$100, exceeding that used in our previous work by nearly two
orders of magnitude.  As the PPLN incoupling efficiency is $\approx$~60$~\%$, and the transmission
through all optics after the PPLN waveguide is $\approx$~80$~\%$, the internal conversion efficiency
in the PPLN waveguide is $>$~70$~\%$.

As mentioned above, single InAs QDs embedded in microdisk cavities
are used as true single photon sources for the quantum frequency
conversion experiments. Here, we present results from two microdisk
devices named as M1 and M2. Figure~\ref{fig:Fig2}(a) shows a low
temperature (T\,=\,10\,K) micro-photoluminescence ($\mu$-PL)
spectrum of device M1, which was obtained under cw excitation with
an energy above the GaAs bandgap. Two bright lines seen in the
spectrum are identified as a neutral excitonic emission from a
single QD (970.2\,nm) and a cavity mode emission (969.8\,nm) with a
quality factor Q\,=\,12500. The identification of the lines is done
based on their excitation power dependence, which revealed first a
linear increase with the power and then a clear saturation for the
QD line, while the cavity mode emission dominated the spectrum at
elevated power conditions as expected. For further investigations,
the QD emission line was spectrally filtered by using a volume Bragg
grating whose output was coupled to a single mode fiber.
Figure~\ref{fig:Fig2}(b) shows the filtered spectrum of the QD
emission with almost 60\,$\%$ total transmission. The filtered PL
spectrum was first directed to a time-correlated single-photon
counting setup for time-resolved PL measurements. The
lifetime~\cite{ref:Microdisk_N2} of the emission is estimated as
$T_{fast}\,=\,1.12\,$ns$\,\pm\,0.05$\,ns (inset of
Fig.~\ref{fig:Fig2}(b)), close to the values measured for QDs in
bulk material, and indicates almost no influence of the cavity mode
on the radiative properties of the QD emission due to large spectral
mismatch.

The single-photon nature of the collected PL from the QD is verified
by measuring the second-order correlation function $g^{(2)}(\tau)$.
An auto-correlation measurement was performed on the filtered QD
emission under similar excitation conditions close to the saturation
power, the result of which is shown in Fig.~\ref{fig:Fig2}(c). A
clear photon antibunching dip is observed at $\tau\,=\,0$ delay with
$g^{(2)}(0)\,=\,0.33\,\pm\,0.04\,<\,0.5$, indicating the quantum
nature of the measured emission line~\cite{ref:Microdisk_N3}. Having
demonstrated the true single-photon generation, the filtered PL was
directed to the frequency conversion setup, which was pre-aligned
for the optimum conversion efficiency and signal-to-background ratio
by using a cw tunable laser at the exact QD emission wavelength, as
described before. Figure~\ref{fig:Fig2}(d) shows the result of an
auto-correlation measurement performed on the QD emission line (same excitation conditions
as in Fig.~\ref{fig:Fig2}(c)), now
converted to 600\,nm and directed to a free-space HBT setup as shown in Fig.~\ref{fig:Fig1}(a). The
antibunching dip observed at zero-time delay is measured as
$g^{(2)}(0)\,=\,0.24\,\pm\,0.04\,<\,0.5$ demonstrating that the
converted signal is mainly composed of single-photons, as expected.
Due to the narrow bandwidth of the frequency conversion process
providing an additional spectral filtering, the antibunching value
observed after the conversion is improved.

\begin{figure}[t]
\centerline{\includegraphics[width=8.5 cm]{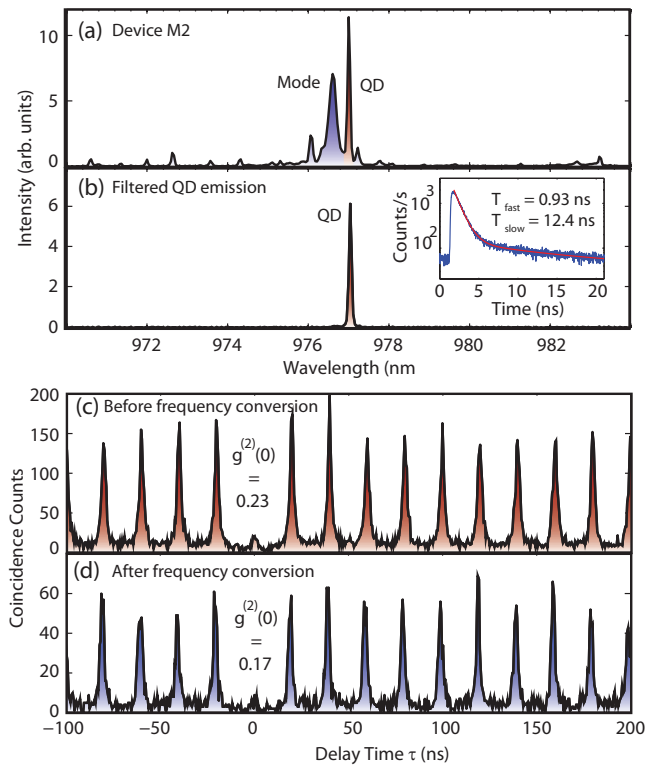}}
\caption{(a) Low-temperature $\mu$-PL spectrum of device M2 under
above-band pulse excitation ($\lambda_{Exc.}\,=\,780$\,nm). A bright
QD emission and a cavity mode emission with a quality factor
Q\,=\,4300 are visible around 977\,nm. (b) Spectrum of QD emission
filtered by a volume Bragg grating. (c)-(d) Second-order
autocorrelation function measurements performed on the QD emission
line before and after the frequency conversion resulted in
$g^{(2)}(0)\,=\,0.23\,\pm\,0.04$ and
$g^{(2)}(0)\,=\,0.17\,\pm\,0.03$, respectively. inset: Time-resolved
PL of the QD emission revealed
$T_{fast}\,=\,0.93\,$ns$\,\pm\,0.05$\,ns and
$T_{slow}\,=\,12.4\,$ns$\,\pm\,0.1$\,ns.} \label{fig:Fig3}
\end{figure}

Similar measurements were carried out under pulsed excitation
conditions on device M2.  Pulsed measurements provide a convenient way to judge the background levels in the
the conversion process, as any noise due to the strong 1550 nm cw beam
will uniformly increase the coincidence rates at all times~\cite{ref:Rakher_NPhot_2010}, whereas the QD
emission is peaked at times corresponding to the repetition period of the excitation
laser. They also better represent how such a system might be used in applications for which
triggered single photon emission is desirable. The PL spectrum for device M2 is shown in Fig. 3(a), where a bright single QD emission line
at 977.04\,nm is visible next to a cavity mode at 976.65\,nm (Q\,=\,4300). Figure~\ref{fig:Fig3}(b)
shows the filtered QD emission and the result of time-resolved PL
measurement as an inset. The lifetime of the QD emission is
estimated as $T_{fast}\,=\,0.93\,$ns$\,\pm\,0.05$\,ns, indicating
again almost no influence of the cavity mode. Before performing
frequency conversion, the emission was directed to the HBT setup for
photon correlation measurements. A result of such an experiment is
shown in Fig.~\ref{fig:Fig3}(c), where the QD was excited with a
pump power close to its saturation. A strong suppression of the peak
at zero time delay to a value of
$g^{(2)}(0)\,=\,0.23\,\pm\,0.04\,<\,0.5$ is observed, thus proving
that the filtered PL line is emitted from a single
QD~\cite{ref:Microdisk_N4}. An auto-correlation measurement was then
performed on the QD emission after it was
converted to 600 nm. As shown in Fig.~\ref{fig:Fig3}(d), the
single-photon nature of the QD emission was preserved during the
conversion process, proven by the value of
$g^{(2)}(0)\,=\,0.17\,\pm\,0.03$. The slightly better value of
$g^{(2)}(0)$ is a result of the additional filtering provided by the
conversion setup. We note that the relatively low coincidence counts
observed in Fig.~\ref{fig:Fig3}(d) are due to the use of a single
mode fiber coupled HBT setup for this experiment, where the single
mode fiber in-coupling efficiency was 30\,\%.

In summary, we have demonstrated quantum frequency conversion of
single photons emitted from a quantum dot.  Photons at 980\,nm are
converted to 600\,nm with a signal-to-background ratio larger than
100 and external conversion efficiency above 40\,\%. Straightforward
engineering of the PPLN waveguide may allow for interfacing
commonly-studied 900~nm to 980~nm band quantum dot single photon
sources~\cite{ref:Michler,ref:Santori2,ref:Moreau,ref:Shields_NPhot}
with quantum memories near 600\,nm, such as in rare-earth doped
crystals~\cite{ref:Timoney} and neutral atoms~\cite{ref:Liu_Chien}.

We thank Edward Flagg for information on volume Bragg gratings and
Matthew Rakher, Lijun Ma, and Xiao Tang for helpful discussions. S.A. and I.A.
acknowledge support under the Cooperative Research Agreement between
the University of Maryland and NIST-CNST, Award 70NANB10H193.


\begin{thebibliography}{31}
\bibitem [{ref({\natexlab{a}})}]{ref:Microdisk_N1}%
  \BibitemOpen
  \href@noop {} {} \bibinfo {note} {The error bars in (c)
  are one standard deviation values and are due to fluctuations in the detected
  count rate on the SPAD.}\BibitemShut {Stop}%
\bibitem [{ref({\natexlab{b}})}]{ref:Microdisk_N2}%
  \BibitemOpen
  \href@noop {} {} \bibinfo {note} {Lifetime data is fit to
  a biexponential function with decay rates $T_{fast}$ and $T_{slow}$, with the
  uncertainties given by the fit and representing one standard deviation
  values.}\BibitemShut {Stop}%
\bibitem [{ref({\natexlab{c}})}]{ref:Microdisk_N3}%
  \BibitemOpen
  \href@noop {} {}  \bibinfo {note} {$g^{(2)}(\tau)$ under cw
  excitation is normalized to the coincidence rate at long time delays
  ($\tau>500$~ns). The uncertainty in $g^{(2)}(0)$ is due to the fluctuation in
  coincidence rate for $\tau>500$~ns, and is the one standard deviation
  value.}\BibitemShut {Stop}%
\bibitem [{ref({\natexlab{d}})}]{ref:Microdisk_N4}%
  \BibitemOpen
  \href@noop {} {} \bibinfo {note} {The $g^{(2)}(0)$ value
  is determined by comparing the integrated area of the peak around time zero
  to the average area of the peaks away from time zero. The uncertainty on this
  value is given by the standard deviation in the area of the peaks away from
  time zero.}\BibitemShut {Stop}%
\end{thebibliography}

%

\end{document}